\documentclass[12pt]{iopart}

\usepackage{bm}%
\usepackage{graphics}%
\usepackage{graphicx}%
\usepackage{epsfig}%
\usepackage{subfig}
\usepackage{multirow}
\usepackage{float}
\usepackage{dcolumn}
\usepackage{upgreek}
\usepackage{cite}
\usepackage{setspace}
\begin{document}

\title[]{Independently tunable dual-spectral electromagnetically induced transparency in a terahertz metal-graphene metamaterial}

\author{Tingting Liu$^{1}$, Huaixing Wang$^{1}$, Yong Liu$^{1}$, Longsheng Xiao$^{1}$, Chaobiao Zhou$^{2}$, Yuebo Liu$^{3}$, Chen Xu$^{4}$ and Shuyuan Xiao$^{1,2,*}$}

\address{$^{1}$~Laboratory of Millimeter Wave and Terahertz Technology, School of Physics and Electronics Information, Hubei University of Education, Wuhan 430205, People's Republic of China}
\address{$^{2}$~Wuhan National Laboratory for Optoelectronics, Huazhong University of Science and Technology, Wuhan 430074, People's Republic of China}
\address{$^{3}$~School of Electronics and Information Technology, Sun Yat-sen University, Guangzhou 510006, People's Republic of China}
\address{$^{4}$~Department of Physics, New Mexico State University, Las Cruces 88001, United State of America}
\ead{syxiao@hust.edu.cn}
\vspace{10pt}
\begin{indented}
\item[]July 2018
\end{indented}

\begin{abstract}
We theoretically investigate the interaction between the conductive graphene layer with the dual-spectral electromagnetically induced transparency (EIT) metamaterial and achieve independent amplitude modulation of the transmission peaks in terahertz (THz) regime. The dual-spectral EIT resonance results from the strong near field coupling effects between the bright cut wire resonator (CWR) in the middle and two dark double-split ring resonators (DSRRs) on the two sides. By integrating monolayer graphene under the dark mode resonators, the two transmission peaks of the EIT resonance can exhibit independent amplitude modulation via shifting the Fermi level of the corresponding graphene layer. The physical mechanism of the modulation can be attributed to the variation of damping factors of the dark mode resonators arising from the tunable conductivity of graphene. This work shows great prospects in designing multiple-spectral THz functional devices with highly flexible tunability and implies promising applications in multi-channel selective switching, modulation and slow light.
\end{abstract}

%
%
%
%
%

\section{Introduction}\label{sec1}
Electromagnetic induced transparency (EIT), as a result of the destructive quantum interference between different excitation pathways, exhibits a sharp transmission window within the broad absorption band in a three-level atomic system\cite{harris1997electromagnetically}. Accompanied with the enhanced transmission, the substantial modification of the dispersion profile enables a significant reduction in the group velocity of light, which can be used in optical buffering and storage. Because the applications of the EIT are severely restricted by the material options and experimental conditions, considerable attention has been paid to metamaterial analogues of EIT in classical optical systems which mimic EIT response via near field coupling between bright and dark mode resonators in artificially tailored nanostructures\cite{zhang2008plasmon, liu2009plasmonic, singh2009coupling, zhu2013broadband, han2014engineering, han2016observation, tian2017low}. Comparing with the EIT effect at a single resonance, the EIT metamaterial with multiple-spectral response shows the capability to manipulation light-matter interaction at multiple frequency regimes and provides new possibilities in optical information processing\cite{artar2011multispectral, zhu2012multi, liu2013multispectral, miyata2014multi, hu2015tailoring, li2016dual, qin2018multispectral, liu2018tailoring}. In particular, the flexibilities in multiple-spectral EIT metamaterial would be highly enhanced when independent and precise manipulation of the multiple resonance responses is obtained.

Recently, continuous interests have focused on the independent modulation of the multiple-spectral EIT resonance. In most of the previous reports, the independent tunability was realized by changing the geometry parameters in a passive way, which sets obstacles for flexibility in practical applications. The advent of graphene with dynamical optical and electric properties facilitates the development of the active metamaterial, especially in terahertz (THz) regime\cite{grigorenko2012graphene, low2014graphene, he2015tunable, he2016further, dadoenkova2017surface, chen2018multiple}. With the doped and patterned processing, graphene can support the strong surface plasmon resonance, and therefore provide an excellent platform to achieve actively tunable single- and multiple-spectral EIT resonance via the controllable near field coupling effects\cite{ding2014tuneable, he2016terahertz, xia2016dynamically, fu2016dynamically, lu2016polarization, zhao2016graphene, yao2016dynamically, he2017implementation, he2018graphene, xia2018plasmonically}. Nevertheless, both the nanofabrication and electrical tuning for these discrete graphene resonators are quite complicated, which hinders the implementation in the functional devices. On the other hand, the manipulation of the doping level of graphene would change the intrinsic resonance frequency of the graphene resonators, therefore the frequency rather than the amplitude modulation of the EIT resonance would be achieved, which is likely to introduce additional noise in the modulation. By integrating monolayer graphene with metal metamaterial arrays, some research works have demonstrated the active modulation of EIT resonance using the plasmonic response of graphene in the mid-infrared, settling the former problem to a great degree\cite{mousavi2013inductive, dabidian2015electrical, sun2017independently, dong2017tunable}. However, these works still focused on the modulation of the resonance frequency rather than the amplitude of the EIT resonance. Most recently, the strong interaction between graphene and THz metal metamaterial has been unveiled\cite{li2016monolayer, xiao2017strong, chen2017study}. Our group and Zhang's group demonstrated the function of the monolayer graphene as a conductive layer in a hybrid metal-graphene metamaterial, which can be employed for the modulation of the resonance amplitude of the single-spectral EIT resonance while maintaining the resonance frequency\cite{xiao2018active, kim2018electrically}. However, the independent and precise modulation on the resonance strength of the multiple-spectral EIT is yet to be investigated.

In this paper, we demonstrate a novel approach based on a hybrid metal-graphene metamaterial to realize the independent modulation of the dual-spectral EIT resonance amplitude at THz frequencies. In the metal structure, the strong near field coupling between a bright cut wire resonator (CWR) in the middle and two asymmetric dark double-split ring resonators (DSRRs) on the two sides gives rise to the dual-spectral EIT resonance with two distinct transmission peaks. By integrating monolayer graphene under the two DSRRs, the independently on-to-off modulation of the two transmission peaks can be obtained through shifting the Fermi level of the corresponding graphene layer. To explain the physical mechanism, we employ the classical three-oscillator model and the distributions of electric field and surface charge density. The active modulation for individual transmission peak can be attributed to the increasing damping factors of the dark DSRRs arising from the conductive effect of graphene layer during corresponding modulation process.

\section{The geometric structure and numerical model}\label{sec2}
The schematic of the hybrid metal-graphene metamaterial is depicted in Figure 1. In this structure, the classical elements including CWR and DSRRs are periodically patterned on the dielectric substrate, and the monolayer graphene is positioned under the bottom of the DSRRs. The CWR and the DSRRs are both made of aluminum (Al) with a thickness of 200 nm, and the common silicon (Si) is adopted for the dielectric substrate. The periods of the unit cell in the $x$ and $y$ directions are $P_{x}=200$ $\upmu$m and $P_{y}=120$ $\upmu$m, respectively. The CWR is located in the middle of the unit cell with the length of $H=100$ $\upmu$m and the width of $W=8$ $\upmu$m. The two DSRRs at the two sides of the CWR have different sizes as the asymmetry is the prerequisite for the dual-spectral EIT resonance. The DSRR on the left side is marked as DSRR1 with the base length of $L_{1}=72$ $\upmu$m and the side length of $D_{1}=68$ $\upmu$m, while the DSRR on the right side is denoted by DSRR2 with the base length of $L_{2}=56$ $\upmu$m and the side length of $D_{2}=56$ $\upmu$m. Both DSRRs share the same width of $W=8$ $\upmu$m and the gap length of $g=12$ $\upmu$m. To obtain the strong near field coupling effects between the resonators, the two DSRRs are placed in close proximity to the CWR with the distance of $s=4$ $\upmu$m.
\begin{figure}[htbp]
\centering
\includegraphics[scale=0.60]{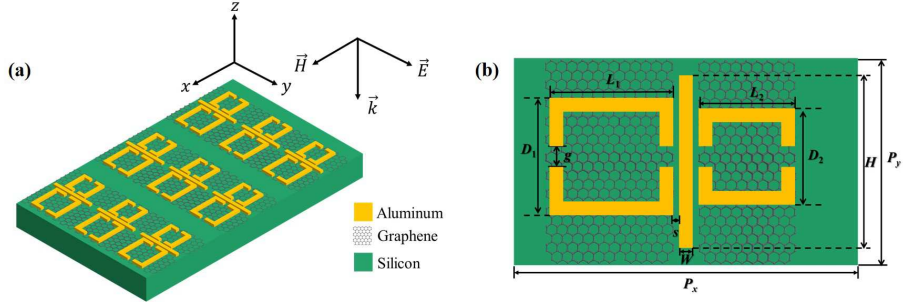}
\caption{\label{fig:1}(a) The structural schematic of the proposed hybrid metal-graphene metamaterial and the incident light polarization configuration; (b) the unit cell of the proposed metamaterial, $P_{x}=200$ $\upmu$m, $P_{y}=120$ $\upmu$m, $H=100$ $\upmu$m, $W=8$ $\upmu$m, $L_{1}=72$ $\upmu$m, $L_{2}=56$ $\upmu$m, $D_{1}=68$ $\upmu$m, $D_{2}=56$ $\upmu$m, $g=12$ $\upmu$m, $s=4$ $\upmu$m.}
\end{figure}

To explore the EIT resonance characteristics of the proposed hybrid metamaterial, the optical response and field distributions are numerically calculated based on the finite difference time domain (FDTD) method. The plane waves are normally incident onto the metamaterial structure along the $-z$ direction and the electronic field is along the $y$ direction, as shown in Figure 1(a). In simulations, the refractive index of the Si substrate is taken as $n_{Si}=3.42$ and it is assumed to be semi-infinite. The frequency-dependent permittivity of Al is calculated by the Drude model
\begin{equation}
    \varepsilon_{Al}=\varepsilon_{\infty}-\frac{\omega_{p}^{2}}{\omega^{2}+i\omega\gamma},\label{eq1}
\end{equation}
where $\varepsilon_{\infty}$ is the dielectric constant at the infinite frequency, $w_{p}$ is the plasmon frequency, $\gamma$ is the damping constant, $\omega$ is the angular frequency of the incident light. Here the parameter values of $\varepsilon_{\infty}=3.7$, $\omega_{p}=2.24\times 10^{16}$ rad/s and $\gamma=1.22\times 10^{14}$ rad/s are adopted for the simulations\cite{ordal1985optical}. The conductivity of graphene is dominated by the contribution of the intraband process in THz regime, and can be derived by the Drude-like model\cite{zhang2015towards, zhao2016tunable}
\begin{equation}
    \sigma_{g}=\frac{e^{2}E_{F}}{\pi\hbar^{2}}\frac{i}{\omega+i\tau^{-1}},\label{eq2}
\end{equation}
where $e$ is the electron charge, $E_{F}$ is the Fermi level of graphene, $\hbar$ is the reduced Planck constant, $\tau$ is the carrier relaxation time. The parameter $\tau$ is dependent on the carrier mobility $\mu$, the Fermi level $E_{F}$ and the Fermi velocity $v_{F}$, i.e., $\tau=(\mu E_{F})/(e v_{F}^{2})$\cite{xiao2016tunable, xiao2018spectrally}. The values of $\mu=3000$ cm$^{2}$/V$\cdot$s and $v_{F}=1.1\times 10^{6}$ m/s from experimental measurements are adopted in the simulations\cite{zhang2005experimental, jnawali2013observation}. As Eq. (2) shows, a continuous tunable surface conductivity of graphene can be obtained by shifting the Fermi level, on the basis of which the actively tunable EIT resonance can be realized in the proposed hybrid metamaterial.

\section{Simulation results and discussions}\label{sec3}
To analyze the coupling effects between the CWR and the two DSRRs in the Al-based metamaterial, the optical properties of the CWR array, DSRR1 array, DSRR2 array and the combined structure array are simulated and their transmission spectra are depicted in Figure 2. When the electric field of the incident plane waves is along the $y$ direction, i.e. parallel to the CWR, the isolated CWR array is directly excited and acts as a bright mode, showing a localized surface plasmon resonance in the transmission spectra represented by the blue curve in Figure 2. The CWR can be viewed as a dipole antenna due to the coupling with the incident field. By contrast, the isolated DSRR1 and DSRR2 arrays are not excited by the incident light because of the structural symmetry with respect to the exciting field. The two DSRRs act as two dark modes and their transmission spectra are represented by the red and green curves in Figure 2. When the three resonators are combined in a unit cell of the proposed metamaterial with a close proximity, the strong near field couplings between the bright CWR and the two dark DSRRs lead to the indirect excitation of the dark DSRRs. The destructive interference of the two excitation pathways, including the direct excitation of the bright CWR by the incident light and the indirect excitation of the dark DSRRs via the coupling effects, gives rise to the EIT resonances. Moreover, the asymmetric structure of the unit cell consisting of the two different DSRRs causes two transmission peaks at different frequencies of 0.39 THz and 0.55 THz. Thus, the dual-spectral EIT resonance can be observed by the orange curve in Figure 2.
\begin{figure}[htbp]
\centering
\includegraphics[scale=0.40]{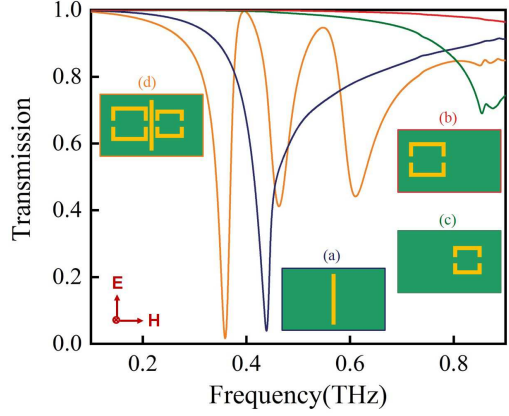}
\caption{\label{fig:2}Transmission spectra of the isolated CWR array (blue curve), the isolated DSRR1 array (red curve), the isolated DSRR2 array (green curve) and the combined array (orange curve) under y-polarized electric field.}
\end{figure}

Next we discuss the independently tunable dual-spectral EIT resonance of the hybrid metamaterial without varying the geometry parameters. In the case without graphene, the two distinct transmission peaks with the resonance amplitude of 99.67$\%$ and 94.75$\%$ at the resonance frequencies of 0.39 THz and 0.55 THz are observed, and called as I and II in the following. When the monolayer graphene is positioned under the dark DSRR1 of the hybrid metamaterial, the active modulation of the resonance amplitude of the transmission peak I at the low frequency can be realized via shifting the Fermi level of graphene, and the variations of the simulated transmission spectra are shown in Figure 3 (a). It can be observed that the amplitude of the transmission peak I successively decreases as the Fermi level of graphene increases. In detail, the transmission peaks experience a sharp decline from 81.49$\%$ to 55.33$\%$ by increasing the Fermi level from 0.25 eV to 0. 50 eV. As Fermi level further increases, the transmission peak I gradually vanishes. At the maximum Fermi level of 1.00 eV, the transmission peak I completely disappears with the amplitude as low as 21.48$\%$ in the spectrum. An on-to-off switching modulation of the transmission peak I in the spectra is obtained in the proposed hybrid metamaterial.
\begin{figure}[htbp]
\centering
\includegraphics[scale=0.60]{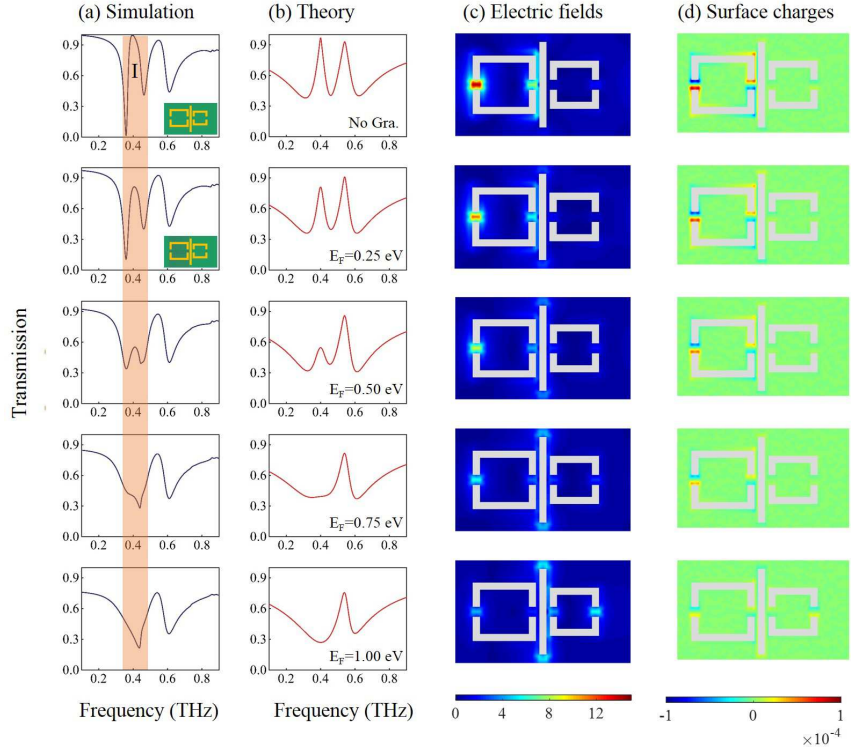}
\caption{\label{fig:3}(a) Simulated transmission spectra and corresponding (b) theoretical fitting results, (c) electric field distributions and (d) surface charge distributions at various Fermi levels of monolayer graphene placed under the DSRR1 in the proposed dual-spectral EIT metamaterial.}
\end{figure}

Similarly, the dynamically tunable transmission peak II at the high frequency can also be achieved via shifting the Fermi level of the monolayer graphene placed under the dark DSRR2. Figure 4(a) illustrates the dependence of the transmission spectra of the proposed hybrid metamaterial on the Fermi level of graphene. The transmission peak II of the dual-spectral EIT resonance displays the similar decreasing tendency with the amplitude varying from 80.01$\%$, 61.37$\%$ to 50.97$\%$ as the Fermi level increases from 0.25 eV, 0.50 eV to 0.75 eV. When the Fermi level reaches the maximum value of 1.00 eV, the transmission peak II is invisible in the spectrum, completing the switch-off of the resonance amplitude at the high frequency. It is noteworthy that the resonance frequencies of the transmission peak I and II remains almost unchanged during the modulation process. Therefore, when the monolayer graphene is respectively placed under the DSRR1 and DSRR2, the transmission peaks I and II can obtain independently modulation through shifting the Fermi level of the corresponding graphene layer. The independent tunability of transmission peaks in the dual-spectral EIT metamaterial shows high efficiency and versatility in practical applications.
\begin{figure}[htbp]
\centering
\includegraphics[scale=0.60]{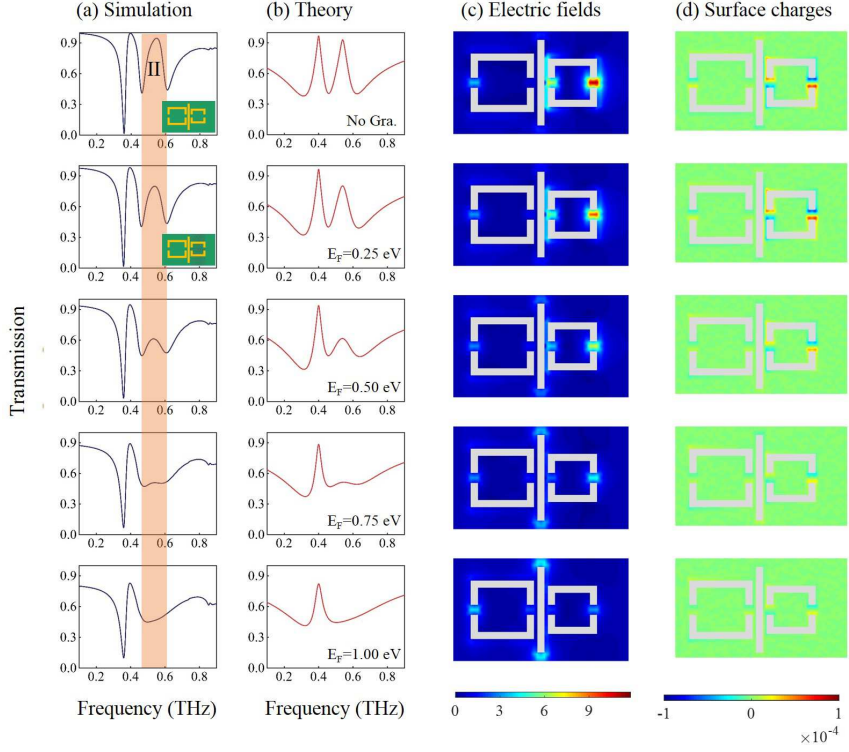}
\caption{\label{fig:4}(a) Simulated transmission spectra and corresponding (b) theoretical fitting results, (c) electric field distributions and (d) surface charge distributions at various Fermi levels of monolayer graphene placed under the DSRR2 in the proposed dual-spectral EIT metamaterial.}
\end{figure}

To explore the modulation mechanism of the dual-spectral EIT resonance, the proposed metamaterial composed of the bright CWR and two dark DSRRs can be quantitatively analyzed by the classical three-oscillator model. The bright CWR which is directly excited by the incident electric field $E$ is represented by oscillator b, and the dark DSRR1 and DSRR2 excited through the near field couplings with bright CWR are represented by oscillators d1 and d2, respectively. According to the coupled differential equations, the energy dissipation of the proposed dual-spectral EIT resonance system is derived as\cite{miyata2014multi}
\begin{equation}
    P(\omega)\propto\frac{1}{A}(\omega_{d1}-\omega-i\frac{\gamma_{d1}}{2})(\omega_{d2}-\omega-i\frac{\gamma_{d2}}{2}),\label{eq3}
\end{equation}
where A is provided by
\begin{eqnarray}
    A&=(\omega_{d1}-\omega-i\frac{\gamma_{d1}}{2})(\omega_{b}-\omega-i\frac{\gamma_{b}}{2})(\omega_{d2}-\omega-i\frac{\gamma_{d2}}{2})\nonumber\\
     &-\frac{\kappa^{2}_{d2}}{4}(\omega_{d1}-\omega-i\frac{\gamma_{d1}}{2})-\frac{\kappa^{2}_{d1}}{4}(\omega_{d2}-\omega-i\frac{\gamma_{d2}}{2}).\label{eq4}
\end{eqnarray}
In the two equations, ($\omega_{b}$, $\gamma_{b}$), ($\omega_{d1}$, $\gamma_{d1}$) and ($\omega_{d2}$, $\gamma_{d2}$) are the resonance frequencies and damping factors of oscillators b, d1 and d2, respectively. $\kappa_{d1}$ and $\kappa_{d2}$ are the coupling coefficients between the oscillators b and d1 and between b and d2, respectively. With the relation between transmission amplitude $T$ and the energy dissipation $P$, i.e., $T=1-P$, the simulated transmission spectra can be analytically fitted using Eq. (3). The theoretical fitting results are depicted in Figure 3 (b) and Figure 4(b), respectively, both of which show reasonable agreement with the corresponding simulated spectra in Figure 3 (a) and Figure 4(a).

For the active modulation of transmission peak I of the dual-spectral EIT resonance, the fitting parameters including $\gamma_{b}$, $\gamma_{d1}$, $\gamma_{d2}$, $\kappa_{d1}$ and $\kappa_{d2}$ as the function of the Fermi level of graphene are presented in Figure 5(a). During the modulation process, $\gamma_{b}$, $\gamma_{d2}$, $\kappa_{d1}$ and $\kappa_{d2}$ remain roughly constant with the increasing Fermi level, while the damping factor $\gamma_{d1}$ of the dark DSRR1 shows a significant increase by more than two orders of magnitude from 0.001 to 0.9 THz. Thus, the actively tunable transmission peak I of the dual-spectral EIT resonance is caused by the variation of the damping factor $\gamma_{d1}$ of the dark DSRR1. In the proposed hybrid metamaterial, the monolayer graphene functions as a conductive layer and bridges the split gap of the DSRR1. With the increasing surface conductivity due to the larger Fermi level, the losses in the dark DSRR1 are greatly enhanced and then the near field coupling effect between the bright CWR and the dark DSRR1 is weakened. Finally, when the Fermi level reaches 1.00 eV, the losses in DSRR1 are too large such that the coupling effect cannot be sustained, leading to the switch-off of the transmission peak I. In Figure 5(b), the fitting parameter values dependent on the Fermi level are provided for the modulation of transmission peak II. It is observed that the parameters $\gamma_{b}$, $\gamma_{d1}$, $\kappa_{d1}$ and $\kappa_{d2}$ stay basically constant while the damping factor $\gamma_{d2}$ of dark DSRR2 increases from 0.011 to 0.41 THz in the modulation process. The similar conclusion can be drawn that the tunable transmission peak II is attributed to the change in $\gamma_{d2}$. This phenomenon can be explained by the same mechanism that the monolayer graphene with increasing surface conductivity induces the reduction of the coupling effect between the bright CWR and the dark DSRR2, and leads to the final disappearance of the transmission peak II.
\begin{figure}[htbp]
\centering
\includegraphics[scale=0.80]{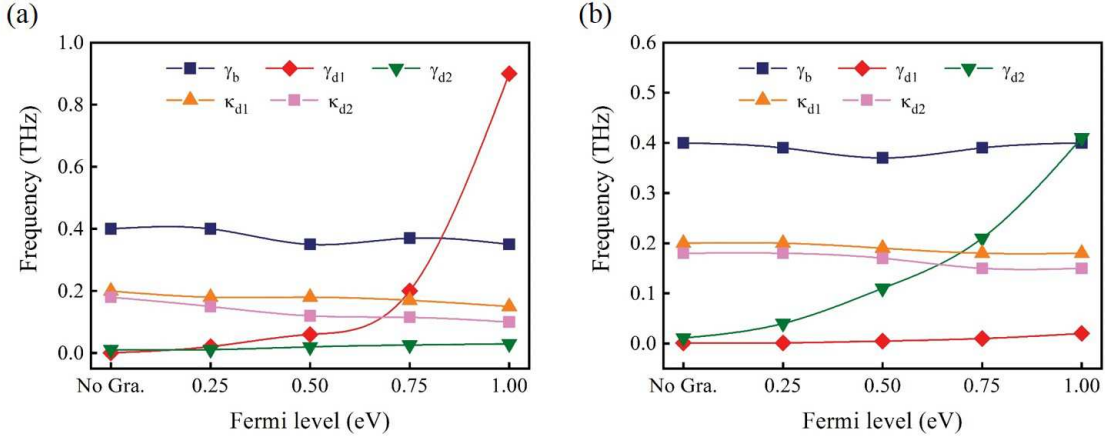}
\caption{\label{fig:5}The variations of the fitting parameters including $\gamma_{b}$, $\gamma_{d1}$, $\gamma_{d2}$, $\kappa_{d1}$ and $\kappa_{d2}$ as the function of the Fermi level of graphene placed (a) under the DSRR1 for the modulation of the transmission peak I; (b) under the DSRR2 for the modulation of the transmission peak II in the proposed dual-spectral EIT metamaterial.}
\end{figure}

To further understand the physical mechanism of the active modulation, the distributions of electric field and surface charge density at the resonance frequencies of the transmission peak I and II are investigated. Figure 3 (c) and (d) show the field distributions at the resonance frequency of the transmission peak I with different Fermi levels of graphene under the DSRR1. For the case without graphene, the strong electric field and the large amount of the accumulated opposite charges can be observed around the ends of the gaps in the DSRR1, while the bright CWR exhibits a quite weak field distribution at the resonance frequency. This result indicates the excitation of the dark DSRR1 through the near field coupling with the bright CWR, which reversely suppresses the excitation of the CWR due to the destructive interference. In this case, the dark DSRR1 shows a nearly zero damping factor and the transmission peak I with high amplitude is observed. As the conductive graphene is placed under the dark DSRR1, the opposite charges around the gaps of DSRR1 are recombined and the electric field is redistributed. With the Fermi level of graphene as 0.50 eV, the initial strong electric field and large charge density around DSRR1 show a remarkable decline, and that around CWR becomes visible. At the maximum Fermi level of 1.00 eV, the graphene with high surface conductivity almost completely recombines and neutralizes the opposite charges around DSRR1. The electric field and the surface charges around DSRR1 are eliminated and that in CWR becomes significant, destroying the destructive interference between CWR and DSRR1 in the EIT metamaterial. Similarly, the distributions of electric field and surface charge density at the resonance frequency of the transmission peak II are plotted in Figure 4 (c) and (d). Without graphene, the strong electric field and surface charge density are visible around the gaps of DSRR2, revealing the indirect excitation of the dark DSRR2 through the coupling effect. Also, it is worthy note that strong field distributions are observed only around DSRR1 at the resonance frequency I and only around DSRR2 at the resonance frequency II, which explains the appearance of the dual-spectral EIT resonance. When the graphene is integrated, the field distributions at the resonance frequency of the transmission peak II in Figure 4 (c) and (d) display similar variations with that in Figure 3 (c) and (d) as the Fermi level increases. The electric field and the surface charge density around the dark DSRR2 gradually weaken due to the recombination effect of the conductive graphene, while that around the bright CWR appear and become significant due to the reduced destructive interference. Consequently, the physical origin of the independently active modulation on the dual-spectral EIT resonance strength essentially comes from the tunable conductivity of the monolayer graphene positioned under the two dark DSRRs.

\section{Conclusion}\label{sec4}
In conclusion, we have provided an approach to achieve the modulation of dual-spectral EIT resonance in a hybrid metal-graphene metamaterial. By integrating the monolayer graphene into a THz metal metamaterial composed of bright CWR and two dark DSRRs, the two transmission peaks of the EIT resonances can be independently tuned via shifting the Fermi level of graphene under the corresponding dark mode resonators. The classical three-oscillator model is employed to describe the near field coupling effects in the proposed metamaterial and the theoretical fitting results agree with the simulated spectra. Combined with the distributions of electric field and surface charge density, the physical mechanism of the active modulation can be attributed to the changing damping factors of the dark mode resonators arising from the conductive effect of monolayer graphene. The proposed hybrid metamaterial provides a new prospective to realize the independently tunable multiple-spectral EIT resonance and shows great prospects in designing THz metadevices with multi-spectral selectivity and highly flexible tunability.

\section*{Acknowledgments}
This work is supported by the National Natural Science Foundation of China (Grant No. 61775064), the Fundamental Research Funds for the Central Universities (HUST: 2016YXMS024) and the Natural Science Foundation of Hubei Province (Grant No. 2015CFB398 and 2015CFB502).

\section*{References}

\providecommand{\newblock}{}

\end{document}